\documentclass[iop]{emulateapj}

\newcommand{\myemail}{\url{hpourima@uark.edu}}
\newcommand {\apgt} {\ {\raise-.5ex\hbox{$\buildrel>\over\sim$}}\ }
\newcommand {\aplt} {\ {\raise-.5ex\hbox{$\buildrel<\over\sim$}}\ }

\shorttitle{Spiral Arm Pitch Angles in Different Wavelength}
\shortauthors{Pour-Imani et al.}

\usepackage{color}
\definecolor{gray}{gray}{0.5}
\usepackage{url}
\usepackage[latin1]{inputenc}
\usepackage{tikz}
\usetikzlibrary{shapes,arrows}
\usepackage{epsfig}
\usepackage{amsmath}
\usepackage{hyperref}

\submitted{Published in The Astrophysical Journal Letters, 827:L2, 2016 August 10}

\begin{document}

\title{Strong Evidence for the Density-wave Theory of Spiral Structure
in Disk Galaxies}

\author{Hamed Pour-Imani\altaffilmark{1,2}, Daniel Kennefick\altaffilmark{1,2},
Julia Kennefick\altaffilmark{1,2},
Benjamin L. Davis\altaffilmark{1,3}, Douglas W. Shields\altaffilmark{1,2}, Mohamed Shameer Abdeen\altaffilmark{1,2}}
\altaffiltext{1}{Arkansas Center for Space and Planetary Sciences, University of Arkansas, 346 1/2 North Arkansas Avenue, Fayetteville, AR 72701, USA; \myemail}
\altaffiltext{2}{Department of Physics, University of Arkansas, 226 Physics Building, 825 West Dickson Street, Fayetteville, AR 72701, USA}
\altaffiltext{3}{Centre for Astrophysics and Supercomputing, Swinburne University of Technology, Hawthorn, Victoria 3122, Australia}

\begin{abstract}
The density-wave theory of galactic spiral-arm structure 
makes a striking prediction that the pitch angle of spiral arms
should vary with the wavelength of the galaxy's image. The reason is that
stars are born in the density
wave but move out of it as they age. 
They move ahead of the density wave inside the co-rotation radius, and 
fall behind outside of it, resulting in a tighter pitch angle at wavelengths
that image stars (optical and near infrared)
than those that are associated with star formation
(far infrared and ultraviolet).
In this study we combined large sample
size with wide range of wavelengths, from the ultraviolet to the
infrared to investigate
this issue. For each galaxy we used an optical 
wavelength image ($B$-band: 445 nm) and images from the \textit{Spitzer Space 
Telescope} at two infrared wavelengths (infrared: 3.6 and 8.0 $\mu$m) and we 
measured the pitch angle with the 2DFFT and Spirality codes
\citep{Davis:2012,Shields:2015}. We find that the $B$-band  and 3.6 $\mu$m
images have smaller 
pitch angles than the infrared 8.0 $\mu$m image in all cases, in agreement 
with the prediction of density-wave theory. We also used images in the
ultraviolet from \textit{Galaxy Evolution Explorer}, whose pitch angles agreed with the measurements
made at 8 $\mu$m. Because stars imaged at those wavelengths have not had time during their short lives to move out of the star-forming region.

\end{abstract}

\keywords{galaxies: evolution --- galaxies: spiral --- galaxies: structure  --- galaxies: fundamental parameters}

\section{Introduction}

Spiral arm structure can serve as an indicator for several properties of 
galaxies including central bulge mass and disk surface density 
\citep{Davis:2015} and thus indirectly central black hole mass 
galaxies including central bulge mass and disk surface density 
\citep{Davis:2015} and thus indirectly central black hole mass 
\citep{Seigar:2008} as well as rotation shear \citep{Seigar:2006}, 
rotational velocity \citep{Savchenko:2011} and weakly, bulge-to-disk 
ratio \citep{Kennicutt:1981}. Evolution in spiral structure can provide clues 
about the evolution of the aforementioned properties.

The density-wave theory of spiral structure in disk galaxies was proposed in 
the mid 1960s by C.C. Lin and Frank Shu \citep{Lin:Shu:1964,Bertin:Lin:1996,Shu:2016}. Their theory envisaged 
long-lived quasi-stationary density waves (also called heavy sound), which 
impose a semi-permanent spiral pattern on the face of the galactic disk.
All subsequent versions
of the theory agree that the density wave causes star formation to occur
by compressing clouds of gas as they pass through the spiral arm.

The brightest stars created in this burst of star formation do not live long enough to travel
far from the position of the spiral density waves and so the eye, when
observing the galaxy in optical wavelengths, picks out the spiral pattern
quite easily. 


\begin{figure*}
\includegraphics[width=18.5cm]{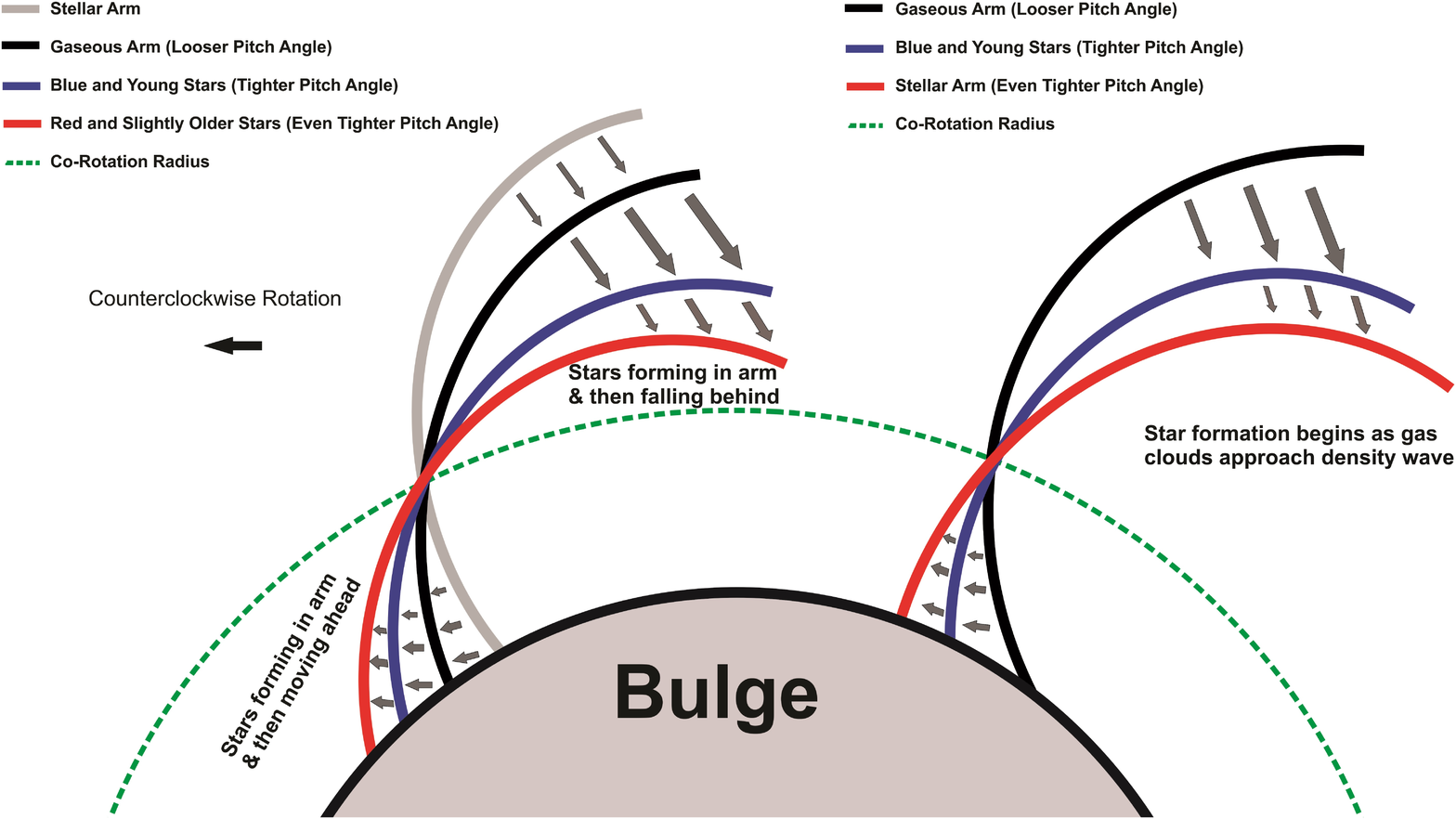}
\caption{Predictions of density-wave theory for spiral-arm structure 
with old stars, blue stars, gas, and dust. 
On the left is a scenario where star formation occurs after 
gas clouds pass through the minimum of the potential of the density wave. 
On the right is a scenario in which star formation occurs as the gas 
clouds approach this minimum of the potential.}
\label{fig:1}
\end{figure*}

The spiral arms predicted by this theory, and those actually observed in
disk galaxies, are approximately
logarithmic spirals. Logarithmic spirals are characterized
by a constant pitch angle 
that has been proposed as a quantifiable feature suitable for
theory testing between
different explanations for galactic spiral structure
\citep{Athanassoula:2010,Martinez-Garcia:2012,Davis:2015}.

Because the spiral pattern moves as if it was a rigid pattern, it follows
that the newly born stars, which are subject to differential rotation,
will quickly move out of the spiral arm.
In fact, since newly born stars are typically obscured from view by the
warm dust-filled clouds associated with star formation (except for the
very brightest UV stars), most of the new stars will be observed only 
when they
leave the spiral arm. In the inner part of the disk, stars move faster
than the spiral-arm pattern and move ahead of the density wave, while in the
outer part of the disk, they fall behind(see Fig. \ref{fig:1}). In between is the co-rotation radius
where stars and the spiral arm move together.\footnote{
In practice, there are
galaxies where the co-rotation radius is quite outside the region
where we measure pitch angle, but the relation between pitch angles 
will still be as described.} It thus follows that the
pitch angle of the pattern formed by the newly formed stars (the bluer stars), 
seen in the optical, is smaller than
the pitch angle of spiral shape formed by the actual star formation region
(we will refer to this region as the gaseous arm), seen in
the far-infrared (which is sensitive to light from the warmed dust of the 
star-forming region itself). In other words, the spiral pattern formed
by the newly formed stars is tighter than the one formed by the gaseous arm
or star-forming region (see either side of 
Figure \ref{fig:1}).

The density-wave
theory predicts that the pitch angle of galactic spiral arms should vary with the
wavelength in which the spiral pattern is observed. This is
in contrast to its main rival, the manifold
theory of spiral structure, which posits that the spiral arms are the result
of highly eccentric orbits of young stars (formed at the end of the galaxy's
bars) that confine the stars to motion along manifolds - tubes running 
across the disk. A key aspect of this theory is
that the pitch angle should not vary with wavelength \citep{Athanassoula:2010}.

This is an excellent opportunity for theory testing and indeed several
attempts have been made to do so, with mixed results. Three studies have looked
at large samples using only two wavebands in the optical 
or near-infrared. Two of these studies have declared
that there is no discernible variation in pitch angle considering only these wavelengths
\citep{Seigar:2006,Davis:2012}. Another study by 
\citet{Martinez-Garcia:2012}, drew
the opposite conclusion. 
Though it is noteworthy that the majority of
galaxies measured in that paper was close to, on or even over
the line of equality from the smaller number that showed the
reported trend (see Fig. 11 of \citet{Martinez-Garcia:2012}).
Two studies by \citet{Martinez-Garcia:2014} and \citet{Grosbol:Patsis:1998} measured 
only five galaxies across several wavebands
spanning the optical and extending into the near infrared or near ultraviolet.
They did observe small differences in pitch angle in a majority of
the galaxies they studied. In addition to these studies, there have been
others that have looked for offsets in position between star-forming regions,
stellar arms, and recently formed stars, both in the radio \citep{Egusa:2009,
Louie:2013} and in the optical \citep{Kendall:2008}.
Thus, results to date on this important question have been inconclusive,
though it would have to be said that the vast majority of galaxies studied
have shown no significant difference in pitch angle across optical and
infra-red wavelengths.

This study makes several advances over previous efforts, most importantly,
the much increased range of wavelengths over which measurements are made.
We make use of \textit{Spitzer} images of galaxies taken at 8 and 3.6 $\mu$m. At 8 $\mu$m,
which images warmed dust in clouds where star formation is occurring, 
we are capable of seeing the region of star formation in the gaseous 
spiral arm. The uv images from \textit{Galaxy Evolution Explorer (GALEX)} show the same region, since 
they are sensitive to stars so young and bright that they are seen while
still in their star-forming nurseries. We do indeed find that the 8 $\mu$m
pitch angles agree with those measured in the uv for those galaxies
(a majority) for which \textit{GALEX} images are available (see Fig. \ref{fig:2}).
The near-infrared and $B$-band images are sensitive to starlight.
We find that the pitch angles of these images are consistently tighter
than those measured for the 8.0 $\mu$m and uv images (see Fig. \ref{fig:2}).
It is clear that the $B$-band images are showing young stars that have
recently left the gaseous arm where they were formed.

Another improvement on earlier work is the large sample size. 
The average error in pitch angle in our sample is 2$^{\circ}$.5. 
To increase confidence in our results, we made use of 
two completely independent
methods of measuring pitch angles. One is an established algorithm involving
a 2DFFT \citep{Davis:2012} decomposition of the galactic image,
and the other is a new approach that compares
the spiral pattern to templates based upon a spiral coordinate system
\citep{Shields:2015}. We found that the two codes agree well and that our
results are independent of the method of measurement used, but we report
only the 2DFFT measurements in this letter.\footnote We took great care
to eliminate the bar from our measurement annulus, as discussed in
\citet{Davis:2012} and also used a function of Spirality
\citep{Shields:2015} to check that the spiral arms in one image actually
corresponded, in angular terms, to spiral arms in the other wavelength
images of the same galaxy.
 We also made use
of a third check on the results, electronically overlaying synthetic spiral
arms of the measured pitch on the galaxy image to let the 
observer's eye provide a check on the validity of each measurement
(see Fig. \ref{fig:3}).


A prediction of this theory is that the pitch angle of spiral arms for galaxies in blue-light wavelength images should be smaller than when imaged in deep infrared-light. Young (blue) stars born in the spiral arms of the galaxies move ahead of the density wave inside the co-rotation radius, and fall behind outside of it. The co-rotation radius is defined as the radius at which the density wave pattern speed is equal to the local rotation speed of stars (which rotate differentially with radius). This implies that blue stars should form slightly tighter arms than the density wave itself does. It means gas and dust involved in star formation should form looser arms with bigger pitch angles than blue and red stars, and blue stars should form bigger pitch angle than red and old stars (because they are short-lived and have less time in which to move ahead of and fall behind the density wave pattern). So the old stars form tighter arms in galaxies (Figure \ref{fig:1}).


\begin{deluxetable*}{llccclllc}
\tablecolumns{7}
\tablecaption{Sample\label{Sample}}
\tablehead{
\colhead{Galaxy Name } & \colhead{ Type } & \colhead{\textit{P} (IRAC 3.6) } & \colhead{\textit{P} ($B$-band)} & \colhead{ \textit{P} (IRAC 8.0)} & \colhead{\textit{P} (FUV Band)} & \colhead{      Image Source      } \\
\colhead{(1)} & \colhead{(2)} & \colhead{(3)} & \colhead{(4)} & \colhead{(5)} & \colhead{(6)} & \colhead{(7)}
}
\startdata
NGC 0157 & SABb & $3.58\pm0.13$ & $8.66\pm0.89$ & $9.32\pm1.01$ &   ...   & IRAC3.60, INT4400, IRAC8.0 \\		
NGC 0289 & SBbc & $9.89\pm1.20$ & $19.71\pm1.94$ & $23.36\pm2.61$ &...& IRAC3.6, CTIO4400, IRAC8.0 \\		
NGC 0613 & SBbc & $19.27\pm2.22$ & $21.57\pm1.76$ & $25.67\pm2.30$ & ...& IRAC3.6, ESO4400, IRAC8.0 \\		
NGC 0628 & Sac & $9.58\pm0.60$ & $9.20\pm0.83$ & $20.60\pm2.28$ & $21.43\pm1.42$ & IRAC3.6, NOT4400, IRAC8.0, GALEX1516A \\
NGC 0925 & SABd & $4.45\pm0.65$ & $7.51\pm3.81$ & $20.10\pm4.69$ & $29.68\pm3.75$ & IRAC3.6, PAL4400, IRAC8.0, GALEX1516A \\		
NGC 1097 & SBb & $6.84\pm0.21$ & $7.54\pm3.49$ & $9.50\pm1.28$ & $16.25\pm2.30$ & IRAC3.6, LCO4400, IRAC8.0, GALEX1516A \\		
NGC 1353 & SBb & $11.14\pm0.70$ & $13.68\pm2.31$ & $17.96\pm1.66$ & ...& IRAC3.6, ESO4400, IRAC8.0 \\		
NGC 1512 & SBab & $4.70\pm2.24$ & $24.80\pm3.43$ & $30.20\pm4.64$ & ...& IRAC3.6, NOT4400, IRAC8.0 \\		
NGC 1566 & SABbc & $15.29\pm2.37$ & $31.20\pm4.80$ & $44.13\pm11.94$ & $45.80\pm2.97$ & IRAC3.6, KPNO4400, IRAC8.0, GALEX1516A  \\		
NGC 2403 & SABc & $12.50\pm1.62$ & $19.35\pm1.57$ & $28.52\pm6.73$ & $23.54\pm0.78$ & IRAC3.6, LCO4400, IRAC8.0, GALEX1516A \\		
NGC 2841 & SAb & $16.13\pm1.63$ & $18.77\pm1.66$ & $22.25\pm2.42$ & $23.26\pm2.31$ & IRAC3.6, LOWE4500, IRAC8.0, GALEX1516A \\
NGC 2915 & SBab & $7.42\pm0.44$ & $7.90\pm0.75$ & $10.40\pm2.08$ & ...& IRAC3.6, KPNO4400, IRAC8.0 \\	
NGC 2976 & Sac & $4.14\pm0.34$ & $5.13\pm0.43$ & $8.36\pm0.40$ & $10.68\pm1.00$ & IRAC3.6, KPNO4400, IRAC8.0, SDSS3551A \\
NGC 3031 & SAab & $15.63\pm6.99$ & $16.19\pm1.23$ & $20.54\pm2.21$ & $20.14\pm 1.90$ & IRAC3.6, JKY4034, IRAC8.0, GALEX1516A \\
NGC 3049 & SBab & $8.60\pm0.46$ & $10.50\pm2.05$ & $16.10\pm2.59$ &...& IRAC3.6, CFHT4400, IRAC8.0 \\		
NGC 3184 & SABcd & $11.92\pm1.77$ & $18.30\pm3.45$ & $23.40\pm3.27$ & $26.75\pm0.55$ & IRAC3.6, KPNO4400, IRAC8.0, GALEX1516A\\		
NGC 3190 & SAap & $16.39\pm2.15$ & $17.67\pm2.34$ & $18.35\pm4.43$ &... & IRAC3.6, CTIO4400, IRAC8.0 \\		
NGC 3198 & SBc & $15.97\pm1.38$ & $18.95\pm2.69$ & $20.59\pm5.95$ & $23.98\pm1.84$ & IRAC3.6, CTIO4400, IRAC8.0, GALEX1516A \\		
NGC 3351 & SBb & $4.60\pm1.92$ & $16.41\pm1.93$ & $22.21\pm6.96$ & $27.17\pm2.11$ & IRAC3.6, CTIO4400, IRAC8.0, GALEX1516A \\	
NGC 3513 & SBc & $19.35\pm2.85$ & $20.27\pm1.67$ & $22.20\pm2.29$ &... & IRAC3.6, CTIO4400, IRAC8.0 \\		
NGC 3521 & SABbc & $16.74\pm1.32$ & $19.28\pm1.92$ & $21.48\pm2.19$ & $24.81\pm2.40$ & IRAC3.6, CTIO4400, IRAC8.0, GALEX1516A \\		
NGC 3621 & SAd & $17.22\pm3.37$ & $18.43\pm3.12$ & $20.81\pm2.72$ & $20.34\pm1.98$ & IRAC3.6, ESO4400, IRAC8.0, GALEX1516A \\
NGC 3627 & SABb & $11.71\pm0.78$ & $16.97\pm1.54$ & $18.59\pm2.85$ & $40.29\pm1.60$  & IRAC3.6, KPNO4400, IRAC8.0, GALEX1516A\\		
NGC 3938 & SAc & $11.46\pm2.32$ & $12.22\pm1.94$ & $19.34\pm3.80$ & $21.45\pm1.87$ & IRAC3.6, LOWE4500, IRAC8.0, GALEX1516A \\
NGC 4050 & SBab & $5.82\pm0.51$ & $6.32\pm1.89$ & $9.00\pm1.01$ &... & IRAC3.6, LCO4050, IRAC8.0 \\		
NGC 4254 & SAc & $28.40\pm4.04$ & $30.01\pm4.36$ & $32.8\pm1.45$ & $38.66\pm3.91$ & IRAC3.6, INT4034, IRAC8.0, GALEX2274A \\
NGC 4321 & SABbc & $18.60\pm1.69$ & $15.06\pm1.20$ & $24.46\pm3.76$ & $28.49\pm1.26$ & IRAC3.6, KPNO4331, IRAC8.0, SWIFT2030A \\		
NGC 4450 & SAab & $12.59\pm2.63$ & $16.62\pm1.45$ & $21.2\pm3.87$ & $22.99\pm5.43$ & IRAC3.6, LOWE4500, IRAC8.0, GALEX2267A \\
NGC 4536 & SABbc & $17.3\pm1.75$ & $33.74\pm4.8$ & $52.22\pm2.42$ & $55.59\pm2.75$ & IRAC3.6, KP4400, IRAC8.0, GALEX1516A\\		
NGC 4569 & SABab & $8.64\pm0.52$ & $19.05\pm2.42$ & $38.55\pm6.44$ & $42.11\pm5.80$ & IRAC3.6, PAL4050, IRAC8.0, GALEX1516A \\		
NGC 4579 & SABb & $11.44\pm0.57$ & $13.8\pm3.25$ & $30.73\pm4.73$ & $33.98\pm3.69$ & IRAC3.6, KPNO4400, IRAC8.0, SWIFT2030A\\	
NGC 4725 & SABab & $3.04\pm1.83$ & $7.40\pm0.35$ & $10.80\pm1.04$ & $13.60\pm1.29$  & IRAC3.6, KPNO400, IRAC8.0, GALEX2267A\\		
NGC 4736 & SAab & $8.19\pm2.94$ & $8.41\pm1.34$ & $14.09\pm5.11$ & $14.98\pm2.31$ & IRAC3.6, PAL4400, IRAC8.0, GALEX1516A \\
NGC 4939 & SAbc & $10.93\pm3.60$ & $11.20\pm1.07$ & $16.25\pm4.94$ & ...& IRAC3.6, CTIO4400, IRAC8.0 \\		
NGC 4995 & SABb & $12.40\pm4.36$ & $13.00\pm2.87$ & $15.90\pm3.66$ & ...& IRAC3.6, LCO4050, IRAC8.0 \\		
NGC 5033 & SAc & $7.09\pm0.46$ & $10.46\pm2.66$ & $13.91\pm4.42$ & ...& IRAC3.6, KPNO4400, IRAC8.0 \\	
NGC 5055 & SAbc & $16.35\pm1.78$ & $19.31\pm1.63$ & $20.63\pm2.11$ & $20.29\pm5.87$ & IRAC3.6, PAL4360, IRAC8.0, GALEX1516A \\
NGC 5474 & SAcd & $12.11\pm1.15$ & $13.84\pm6.22$ & $19.12\pm3.22$ & $19.91\pm2.65$ & IRAC3.6, JKY4034, IRAC8.0, GALEX1516A \\
NGC 5713 & SABb & $12.20\pm0.32$ & $18.76\pm3.10$ & $34.79\pm5.01$ & $27.40\pm1.20$ & IRAC3.6, CTIO4400, IRAC8.0, GALEX1516A \\		
NGC 7331 & Sab & $17.13\pm2.63$ & $20.10\pm1.85$ & $21.65\pm2.15$ & $22.54\pm2.41$ & IRAC3.6, KPNO4400, IRAC8.0, GALEX1516A \\
NGC 7793 & SAd & $10.98\pm1.6$ & $12.16\pm2.1$ & $16.34\pm5.47$ & $16.89\pm1.87$ & IRAC3.6, ESO4400, IRAC8.0, GALEX1516A \\

\enddata
\tablecomments{Columns:
(1) Galaxy name;
(2) Hubble morphological type. 
(3) pitch angle in degrees for infrared 3.6 $\mu$m;
(4) pitch angle in degrees for $B$-band 445 nm;
(5) pitch angle in degrees for infrared 8.0 $\mu$m;
(6) pitch angle in degrees for FUV 1516 $\textrm{\AA}$;
(7) telescope/literature source of imaging .
}
\end{deluxetable*}

\section{Data and Analysis}




Our sample of 41 galaxies is drawn from the \textit{Spitzer} Infrared Nearby Galaxies Survey, which consists of imaging from
the Infrared Array Camera \citep{Fazio:2004}, 
selecting those galaxies with imaging at both 3.6 and 8.0 $\mu$m and that had available optical imaging in the $B$-band (445 nm)
as found in the NASA/IPAC Extragalactic Database (NED; see Table 1 for $B$-band image sources).
Twenty-eight (28) of these galaxies also have available ultraviolet imaging from archived  
\textit{GALEX} data at two wavelengths,
far-UV (FUV) 1350-1780 $\textrm{\AA}$ and near-UV (NUV) 1770-2730 
$\textrm{\AA}$, as indicated in Table 1.

\section{Results}

Spiral-arm pitch angles of our sample galaxies were measured in three or four
wavelength bands using the 2DFFT code \citep{Davis:2012} and checked with the
Spirality code of \citet{Shields:2015}(not reported here).

The $B$-band images are sensitive primarily to newly born stars
that have emerged from their stellar nurseries. For the images
at 3.6 $\mu$m it is expected that older stars dominate.  By contrast, 
at 8.0$\mu$m, we can see details of gas and dust in 
spiral arms \citep{Elmegreen:2011}, as this waveband 
is sensitive to dust warmed by nearby star formation. Finally, the \textit{GALEX}
images at 1516 $\textrm{\AA}$ are sensitive to the brightest O-type stars 
with the shortest lives, visible while still in the star-forming region.

\begin{figure*}
\includegraphics[width=10 cm]{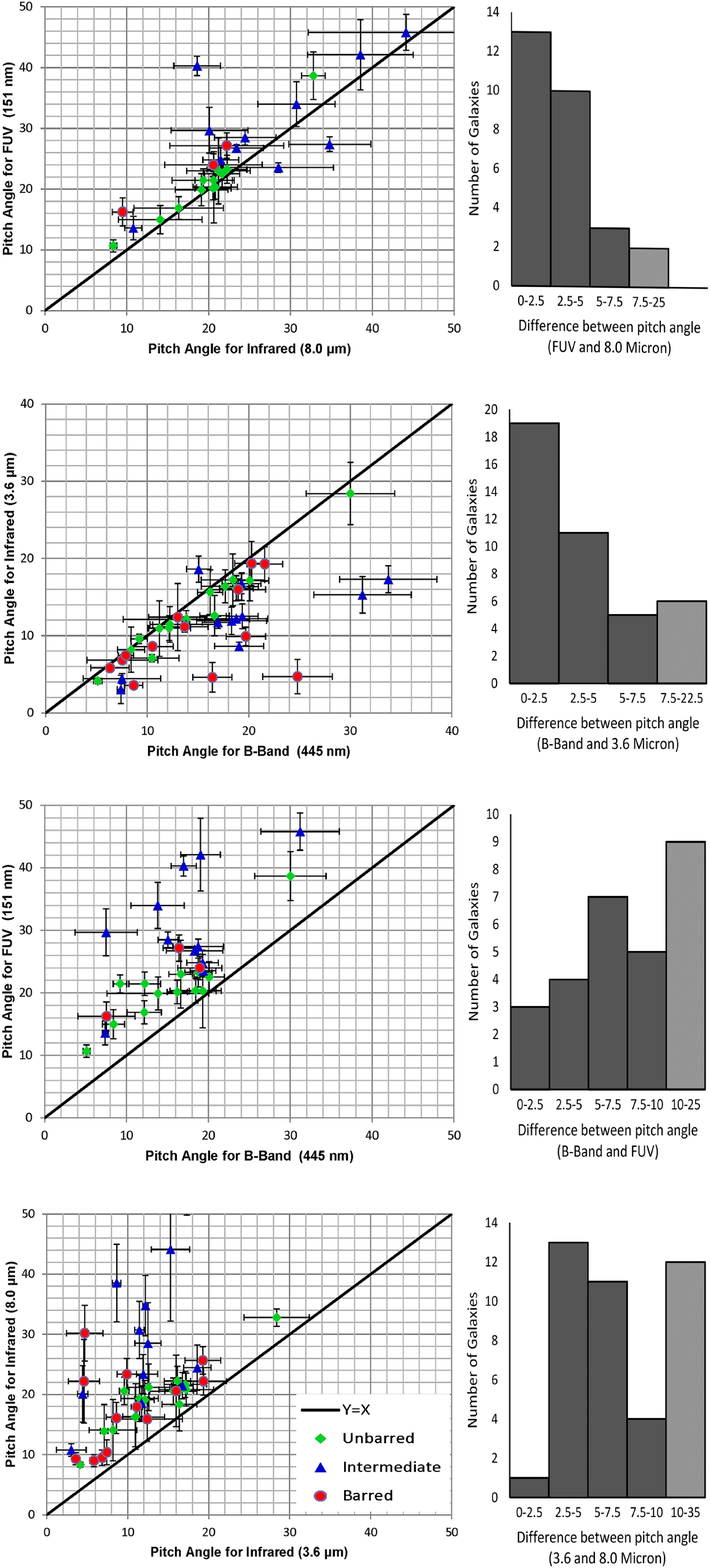}
\centering
\caption{Comparisons between pitch angles measured at different
wavelengths. Each point on the plots represents an individual galaxy
positioned according to the measurement of its spiral-arm pitch angle
at two different wavelengths. The histograms show the distribution
of pitch angle differences in terms of the number of galaxies found
in each bin. The histograms for the top plot shows that the
8.0 $\mu$m and far ultraviolet (FUV) wavelengths are fundamentally equal
since the greatest number of galaxies have pitch angles at these
wavelengths that agree to better than 2$^{\circ}$.5. The same is true
for the second plot, comparing $B$-band with 3.6 $\mu$m infrared images.
In contrast, we can see that $B$-band and FUV pitch angles (third plot)
and 3.6 and 8 $\mu$m pitch angles (bottom plot) are different
from each other, since in both cases the greatest number of galaxies
have a pitch angle difference of between 2$^{\circ}$.5 and 5$^{\circ}$ (see
the relevant histograms), with very
few found below 2$^{\circ}$.5. Images of 41 galaxies were used at
445nm,151 nm, 3.6, and 8.0 $\mu$m and 28 of these also had images at
FUV (151 nm).}
\label{fig:2}
\end{figure*}
\begin{figure*}
\includegraphics[width=18.5cm]{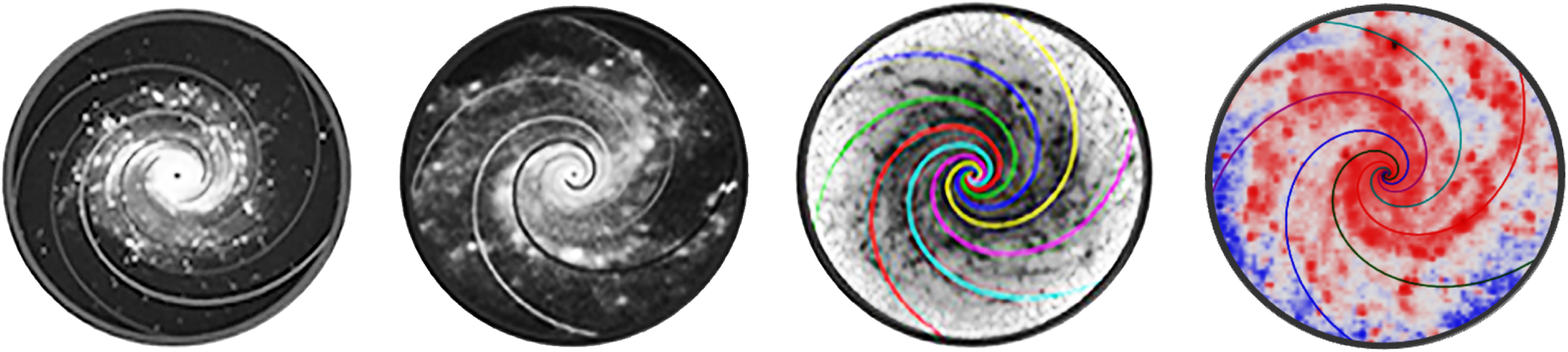}
\caption{Pitch angle for NGC 3184  with different wavelengths. Left to right: -11$^{\circ}$.92 (3.6 $\mu$m), -18$^{\circ}$.30 ($B$-band), -23$^{\circ}$.40 (8.0 $\mu$m), -26$^{\circ}$.75 (FUV).}
\label{fig:3}
\end{figure*}

Our results for 28 local galaxies show that spiral 
arms for images at 8.0$\mu$m are clearly very similar in
pitch angle to the same spiral arms observed in the far-UV by \textit{GALEX}
(see Fig. \ref{fig:2}, top panel). If
they are different, as the histogram in Fig. \ref{fig:2} suggests, it is by less
than 2$^{\circ}$.5 of pitch in most cases, which is the average error in
our measurements. 
Similarly, as seen in the second panel of Fig. \ref{fig:2}, the pitch angles
of 41 galaxies (the entire sample)
in the $B$-band and 3.6 $\mu$m images are also close to the line of
equality, clustered largely within 2$^{\circ}$.5 of it, as the histogram
shows. 

By contrast, the third panel shows that the $B$-band and Far-UV images
clearly disagree in pitch
angle. The $B$-band images have consistently tighter pitch angles,
with the histogram showing that they typically differ by 5$^{\circ}$ or more.
Similarly, there is also a clear difference,
in the bottom panel, between the pitch angles in the two infrared
bands. Once again it is the stellar waveband (3.6 $\mu$m) that is
consistently tighter in pitch angle than that associated with the
star-forming region (8.0 $\mu$m).



\section{Discussion}
It is apparent from Figure \ref{fig:2} that the far-UV and the 8.0 $\mu$m images have
essentially the same pitch angle. This supports our
argument that these wavebands both image the star-forming region.
It forms a
spiral pattern that is noticeably looser than that formed in both the
$B$-band and the 3.6 $\mu$m images, which both image stars.
Thus, we confirm the
picture from the left-hand side of Figure \ref{fig:1}, 
in which the star-forming region or gaseous arm (UV and 8.0 $\mu$m)
has a larger pitch angle than that formed by the bluer stars ($B$-band)
and the redder stars (3.6 $\mu$m).  The region just downstream from
the spiral arm has just as many old disk stars as any other region of
the disk, but the population has been augmented by recently formed
reddish stars. Thus, even in the red or infrared,
the region associated with newly formed stars is brighter than other
parts of the disk. For several intermediate and barred galaxies, the difference in pitch angle in the different wavebands is very high, which may mean that the 
pitch angle derived is biased by the presence of a bar.

We employed a Monte Carlo technique to generate two-dimensional Gaussians about each data point 
(based upon the associated measurement errors) 
to see what were the chances of finding
counter-examples to our reported trend. We find that 
there is on the order of a 1\% chance of contradicting the claim
of tighter pitch angles for the stellar sources than for the star-forming regions (Figure 2, bottom two panels).
For the cases where the two bands are both sampling stellar sources (Figure 2, second panel) or star-forming regions (Figure 2, top panel), the 
chance of finding a contradictory result are on the order of 10\%.


Our results are compatible with those in \citet{Davis:2012} and 
\citet{Seigar:2006}. Both claimed to see no noticeable change in pitch angle
at wavebands that image stars, Davis {\it et al.} between $B$ and $I$ bands and
Seigar {\it et al.} between $B$ on the one hand and either $K\prime$ or $H$ bands on
the other. We believe this is consistent with our picture that there are
fundamentally two discernible pitch angles, one that images the 
gaseous arm and one that images stars that have moved
out from the star-forming region to form a tighter spiral pattern nearby and
that crosses the gaseous arm at the co-rotation radius. We
do see a modest difference between the bands at the extreme ends of
this range, from $B$-band to 3.6 $\mu$m (see Fig. \ref{fig:2}), but it is small at best.

Another study is \citet{Martinez-Garcia:2012}, which compares pitch angles
in $B$ and $H$ for a good-sized sample of galaxies. Although Martinez-Garcia
sees an overall tendency for the $B$-band pitches to be tighter than the
$H$-band pitches (see Fig. 11 of \citet{Martinez-Garcia:2012}), we note 
(as stated above) that many of his objects are consistent with an equality
between the pitch angles at optical and near-infrared images. 

Our results are not compatible with the claim made by both \citet{Martinez-Garcia:2014} and \citet{Grosbol:Patsis:1998} 
that they see a tendency for pitch angles to be
tighter at blue wavelengths than at red. Broadly speaking, we see the 
opposite, looser spirals in the ultraviolet, growing tighter in the blue,
and perhaps tightening a little further into the red. We find, nevertheless,
that there is some
important common ground between our work and that of \citet{Martinez-Garcia:2014}.

Grosbol and Patsis claim
a difference in pitch angle, for four objects \footnote{Grosbol and Patsis 
measure five
galaxies but only have $B$-band measurements for four of them.}, 
between the $B$-band and the I band, which is tighter in $B$ for all four.
In Fig. \ref{fig:1} we present the results of Davis et al which also measures
$B$-band versus $I$-band for a larger sample of galaxies. The reader will
note that none of Grosbol and Patsis' objects are exceptional in this
sample. They are close to galaxies with a similar difference between 
$B$ and $I$ as measured by Davis et al. But the over all spread of Davis
et al's results straddles the line of equality. Thus we believe that a
larger sample in Grosbol and Patsis would have shown a similar result,
that there is no significant difference in pitch angle between the $B$
and $I$ bands. They claim an even larger difference between $B$ and $K'$,
in disagreement with the much larger sample of Seigar et al, who see
no consistent tendency for $B$ or $K'$ to be tighter than the other. Our
difference between B and 3.6 $\mu$m is the opposite to that claimed
between $B$ and $K'$ by Grosbol and Patsis (they see $B$ as the tighter pitch,
we see it as looser). We do not have high quality images in $K'$ for a
direct comparison. We do note that our measurements in B for two of
their objects agree reasonably well with theirs, so this is not simply
the result of two different methods of measuring pitch angle.


We agree completely with one key result of \citet{Martinez-Garcia:2014}:
that pitch angles in images taken at the $H$$\alpha$
line agree well with the pitch measured in the $u$ band. They argue
that these images are capturing the star-forming region. We agree
since we have measured some of our objects in the $u$ band and find
results in agreement with those given here for the UV and 8.0 $\mu$m
bands. Therefore, it seems likely that both they and we are successfully
imaging the gaseous arm in a number of different widely separated
wavebands. However, they find that images in the $g$, $r$, $i$, and $z$ bands tend
to show tighter pitch angles, compared to $u$ and $H$$\alpha$. So, contrary
to us, they claim that the stellar pitch angles are tighter than the
pitch angle of the gaseous arm where stars are formed. 

Their analysis, which is based on the theoretical work of \citet{Kim:2013}, is that their red bands are imaging the stellar spiral
arm where the density wave causes old disk stars to crowd closer together. 
The pitch angle of this ``stellar arm'' should be the largest 
because the density wave moves in a 
fixed pattern. Everything else moves ahead of the stellar arm
when inside the co-rotation radius and falls behind outside of it, as 
illustrated in Fig. \ref{fig:1} (this is a simplified account, but qualitatively
matches the more complex picture coming from density-wave theory, as
seen in \citet{Kim:2013}. The stellar arm 
is where gas clouds passing through the density wave begin
their collapse (gray arm in Fig. \ref{fig:1}). A short while later their
gravitational collapse has proceeded to the point where they are giving birth 
to stars. By this time they have
moved to a new position that is referred to as the gaseous arm 
(black arm in Fig. \ref{fig:1}). Because 
the stellar arm and the gaseous arm
cross each other at the co-rotation radius, 
\citet{Martinez-Garcia:2014} and \citet{Grosbol:Patsis:1998} 
are looking for a gradient of this type:
red spiral arms tighter than blue. Equivalently, the 
stellar arm is tighter than the gaseous
arm, as depicted in Fig. 7 of \citet{Kim:2013}, which is based upon complex
theoretical modeling of the dynamics within the spiral arm created
by the density wave. 

We interpret our results, in contrast, not as the result of a gradient
across the spiral arm itself, but a gradient produced by migration of
new stars, born inside the spiral arm, as they pass out of it.
As  \citet{Martinez-Garcia:2014} say in their paper, speaking about the stars born 
in the spiral arm, ``these young stars will then gradually age, 
as they leave the place where they were born, and produce a gradient toward
the red in the opposite direction.'' That is to say, when
we image the stars that originate in the star-forming region,
these stars will have moved further on from that region (ahead in the
inner disk, behind in the outer disk) and so the pitch angle gradient
will be, as they say, in the opposite direction: UV stars will have
the loosest pitch, blue stars a tighter pitch. This is, of course,
exactly what we see in our sample. In fact, the reddest
stars in our sample may have the tightest pitch angle of all. We conclude,
therefore, that these are also stars born in the spiral arm that have
migrated out of it. We stress, however, that this issue is not critical
to the confirmation of the density-wave theory. It is sufficient to
note the difference in pitch angle between the newly born blue stars
and the 8.0 $\mu$m and UV images that sample the gaseous spiral arm.

One possible interpretation is that our 3.6 $\mu$m images are not capturing the old stars
in the ``stellar arm'' (gray line in Fig. \ref{fig:1}). 
Since our 3.6 $\mu$m images are, if anything, 
slightly tighter than the $B$-band images,
we might interpret this as evidence that these stars are also 
recently born.  Of course, they mingle with and augment the light from
a population of older disk stars
that themselves just passed through the spiral arm. 
Thus, we are seeing the
``gradient ... in the opposite direction'' referred to in  \citet{Martinez-Garcia:2014}.
Rather
than seeing a color gradient within the spiral arm itself we may be 
seeing
a color gradient created by stars moving downstream from their formation
within the spiral arm.

Another interpretation is possible, however, based upon the notion,
proposed in some versions of density-wave theory \citep{Roberts:1969, Gittins:2004}, that the star formation begins to occur as the gas clouds
approach the density wave (see the right-hand side of Figure \ref{fig:1}). In this
case, the gaseous or star formation arm should have a looser pitch angle
than the stellar arm consisting of old red disk stars concentrated by the
density wave, which is what we see. New blue stars formed in the gaseous
arm move downstream as described earlier and end up close to the position
of the stellar arm. This scenario is clearly compatible with
our results.

We hope that in future work we may decide between these two interpretations
by studying individual galaxies and their dynamics in more detail to determine
which fits better with observations. In this context, it is worth noting that
a few galaxies in Fig. \ref{fig:2}, all barred, have very large changes in pitch 
angle that are hard to reconcile with either scenario. These anomalies
could be due to
difficulties in measurement (pitch angles of these galaxies all have large error bars).
Increasing the sample size may help in identifying the reason for these
odd results.


Regardless of the interpretation,
we find our results to be a strong confirmation of the density-
wave theory. The model that our results support is one in which a star-forming
region rotating with a fixed spiral pattern gives birth to stars that 
move downstream from the spiral arm, with a pitch angle altered by shear
(differential rotation). This is a prediction of the density-wave theory
that is not replicated by its rivals. 
The strongest competitor to the density-wave theory currently
is the manifold theory. As noted by \citet{Athanassoula:2010}, this theory 
finds that the spiral patterns are formed by stars moving along orbital
trajectories within certain elliptical manifolds. It predicts 
that all stars and gas should move together, with no color gradient.
This is contrary to the evidence we present in this letter.

Images from deep infrared wavelength (3.6 and 8.0 $\mu$m) unlike images taken at optical wavelengths show us the spiral arms patterns traced by old stars (near-infrared) and gas and dust (far-infrared). For each galaxy we used an optical wavelength image ($B$-band: 445 nm) and another image from the Spitzer Space Telescope in the deep infrared range and we measured the pitch angle with the 2DFFT code and a completely independent code called Spirality (which uses templates with Fourier transforms to measure pitch angle). Our results for 41 NGC galaxies show that spiral arms for images with optical wavelength 445 nm (more details of blue stars) are clearly tighter than spiral arms in infrared wavelength 8.0 $\mu$m (more details of gas and dust \citep{Elmegreen:2011}) in all cases and spiral arms for images with infrared wavelength 3.6 $\mu$m (more details of red and old stars \citep{Elmegreen:2011}) are clearly tighter than spiral arms in optical wavelength 445 nm (more details of blue stars), in agreement with the prediction of density wave theory (Figure \ref{fig:1}).

\acknowledgments
The authors thank Matthew Bershady and Bret Lehmer for valuable 
suggestions that guided our research. This  research has  made  use of  the 
NASA/IPAC Extragalactic  Database, which  is operated by  
the Jet Propulsion  Laboratory,  California  Institute  of  Technology,  
under contract with the National Aeronautics and Space Administration.

\end{document}